\documentclass[aps,onecolumn,prd,showpacs,nofootinbib]{revtex4}
\usepackage{amsmath}
\usepackage{graphicx}
\usepackage{dcolumn}
\usepackage{bm}
\usepackage{amssymb}
\usepackage{latexsym}

\def\be{\begin{equation}}
\def\ee{\end{equation}}
\def\ba{\begin{eqnarray}}
\def\ea{\end{eqnarray}}

\begin{document}

\title{\Large{DBI-Curvaton} \vspace{6mm}}

\author{Sheng Li$^1$}
\email{lisheng06@mails.gucas.ac.cn}

\author{Yi-Fu Cai$^2$}
\email{caiyf@ihep.ac.cn}

\author{Yun-Song Piao$^1$\vspace{4mm}}

\email{yspiao@gucas.ac.cn}

\vspace{16mm}

\affiliation{$1$ College of Physical Sciences, Graduate School of
Chinese Academy of Sciences, Beijing 100049, China \vspace{1mm}\\
$2$ Institute of High Energy Physics, Chinese Academy of Sciences,
P.O. Box 918-4, Beijing 100039, China}

\begin{abstract}
\vspace{3mm}
\begin{center} \textbf{Abstract}\vspace{3mm}
\end{center}

In this paper, we study a curvaton model described by a
Dirac-Born-Infeld-like action. We calculate the spectrum of
curvature perturbation induced by DBI-curvaton and estimate its
non-Gaussianity. We find that in the limit of low sound speed the
amplitude of curvature perturbation is enhanced more than that in
DBI inflation. This result also means that the inflationary scale
with DBI-curvaton may be lower than that in usual curvaton model.
In addition, we also find that the local non-Gaussianity level in
DBI-curvaton is generally about 10 times larger than that in usual
curvaton model, which is interesting for current observations.
This work in some sense explores a new application of
brane-cosmology in inflation.

\end{abstract}


\maketitle

\section{Introduction}

The curvaton mechanism is an interesting proposal for explaining
the observed scale-invariant primordial density perturbation in
the frame of inflation \cite{Lyth:2001nq}\footnote{Some earlier
works on curvaton mechanism have been considered in Refs.
\cite{Mollerach, Linde:1996gt, Enqvist:2001zp, Moroi:2001ct}.}. In
this scenario, at least two fields are required, in which the
so-called ``curvaton" field is subdominant during inflation but
comes to dominate after inflaton decays. When the curvaton is
subdominant it provides iso-curvature perturbations during
inflation, and afterwards, these iso-curvature perturbations
convert to adiabatic ones when curvaton starts to dominate the
universe. After the curvaton decays, the universe enters the
standard thermal history by curvaton reheating\cite{Feng:2002nb,
Liddle:2003zw}, and this process seeds the adiabatic initial
condition for the perturbations of other species, such as
radiation and matter \cite{Lyth:2002my, Lyth:2003ip}.

In this paper, we study a curvaton model which is described by a
Dirac-Born-Infeld-like (DBI) action. DBI action has a
non-canonical kinetic term, thus its dynamics is different from
the usual field theory. From the calculation, we find that this
type of curvaton model generates iso-curvature perturbations
suppressed by the sound speed $c_s$ which will be introduced in
the context. Moreover, after these modes convert to adiabatic
ones, the curvature perturbation is still enhanced by $c_s$ which
is similar to that in DBI inflation\cite{Silverstein:2003hf,
Chen:2005ad}. However, since the final evolution of a DBI action
is a process of tachyon condensate, this presents a new exit
mechanism for curvaton model which do not need an oscillating
phase. In this case, the primordial curvature perturbation is
determined by the warping factor and the sound speed, but
unrelated to the potential. Further, we extend the perturbations
into second order and simply estimate the local form of the
non-Gaussianity $f_{NL}$, and find that its value is in order
$O(10)$ which is mildly bigger than that in the usual curvaton
model.

The paper is organized as follows. In Section II, we simply review
the standard curvaton model and present the results in this usual
scenario. In Section III, we propose a DBI-curvaton model, then
study the iso-curvature perturbations during inflation and give
its spectral index. In Section IV, we investigate the process for
iso-curvature modes converting to adiabatic, especially discuss
the case of tachyon condensate, and finally present our results.
Section V includes the conclusion and discussions.

In this letter, we take the conventions that a dot represents the
derivative with respect to the cosmic time, a comma denotes the
normal derivative, and the natural units $M_p^2=c=1$ is taken.

\section{The Curvaton Scenario}

To begin with, we would like to review a curvaton model with a
canonical kinetic term.

In Ref. \cite{Lyth:2001nq}, the authors have proposed a new
mechanism to generate the scale-invariant curvature perturbations.
In this mechanism a scalar field, dubbed curvaton, is introduced
which is very light during inflation. This scalar field is
described by
\begin{eqnarray}
\emph{L}_\sigma=\frac{1}{2}{\dot{\sigma}}^2-\frac{1}{2}(\nabla{\sigma})^2-V(\sigma)~,
\end{eqnarray}
and it lives in an inflationary Friedmann-Robertson-Walker
spacetime characterized by
\begin{eqnarray}
ds^2 = - dt^2+a^2(t)\delta_{ij}dx^idx^j~.
\end{eqnarray}
During this era, the hubble parameter $H\equiv\dot a/a$ is nearly
constant, so that we can define a slow-roll parameter
$\epsilon\equiv-\dot H/H^2$ which is almost zero.

In the curvaton mechanism, the curvature (adiabatic) perturbations
are not originated from the fluctuations of the inflaton field,
but instead from those of curvaton. This takes place in two steps.
Firstly, the quantum fluctuations of the curvaton field during
inflation are converted into classical perturbations,
corresponding to iso-curvature perturbations when cosmological
scales leave the horizon. Secondly, after inflation decays, the
energy density of the curvaton field becomes significant, and the
iso-curvature perturbations are converted into sizable curvature
ones. The curvature perturbations become pure when the curvaton
field or the radiation it decays into begins to dominate the
universe.

Due to the basic scenario of curvaton mechanism, it is required
that the curvaton is very light during inflation and the potential
$V(\sigma)$ is required to be very flat $|V_{,\sigma\sigma}| \ll
H^2$ and much lower than the energy scale of inflation $V \ll
3M_p^2H^2$. So the field $\sigma$ is able to roll down slowly and
has nothing to do with inflation. At the same time, the
fluctuations of inflaton field is negligible, so the energy scale
of inflation is relatively lower and the gravitational wave
amplitude is very small. For example, it requires
$H<10^{-5}\epsilon_H^{1/2}M_p$ for slow-roll inflation and
$H<10^{-5}c_{s}^{1/2}\epsilon^{1/2}M_p$ for DBI inflation.
Therefore, the final spectrum of the curvature perturbations is
usually nearly flat and naturally consistent with current
observations.

During inflation, it follows that on super-horizon scales the
Gaussian perturbation of the scalar field $\delta\sigma \simeq
\frac{H_*}{2\pi}$, where $*$ denotes the epoch of horizon exit $k
= a_* H_*$. Moreover, the spectral index is given by
\begin{eqnarray}\label{indexus}
n_{\sigma}-1 = - 2 \epsilon +2 \eta_{\sigma}~,
\end{eqnarray}
in which $\eta_{\sigma} \equiv V_{,\sigma\sigma}/{3H^2}$.

After inflaton decays, the hubble parameter starts to decrease.
Then the curvaton field enters the second stage during which it
starts to oscillate around the bottom of the potential.
Considering the potential $V$ is quadratic,
$V(\sigma)=\frac{1}{2}m^2\sigma^2$, then the oscillations starts
when $H\sim m$. The curvaton energy density is characterized by
the amplitude of the oscillation,
\begin{eqnarray}
\rho_\sigma(\textbf{x}) = \frac{1}{2}m^2\sigma^2(\textbf{x})~,
\end{eqnarray}
and so the density perturbation is given by
\begin{eqnarray}
\delta\equiv\frac{\delta\rho_\sigma}{\rho_\sigma}=\frac{H_*}{\pi\sigma_*}~.
\end{eqnarray}

Therefore, we finally obtain the curvature perturbation before
curvaton decays $H=\Gamma$,
\begin{eqnarray}
\zeta\equiv-H\frac{\delta\rho}{\dot\rho}\simeq r\delta~,
\end{eqnarray}
where $r$ is the fraction of the radiation decayed from inflaton.

To extend the perturbation to second order, we are able to
investigate the features of non-Gaussianity in curvaton scenario.
For simplicity, we only consider the local form. The relative
magnitude of the second order is conventionally specified by a
parameter $f_{NL}$, which is defined by
\begin{eqnarray}\label{fnl}
\zeta\equiv\zeta_g+\frac{3}{5}f_{NL}\zeta_g^2~,
\end{eqnarray}
and in the usual curvaton model its value takes \cite{Lyth:2002my}
\begin{eqnarray}
f_{NL}=\frac{5}{4r}~.
\end{eqnarray}

From current cosmological observations \cite{Komatsu:2008hk}, an
instantaneous decaying of curvaton is favored if it seeds the
primordial curvature perturbations and decays to cold dark matter,
i.e. $r\simeq 1$. In this case, it is usually considered that a
canonical curvaton model can only produce the local
non-Gaussianity with $f_{NL}\sim 1$. In recent years, curvaton has
been studied in a number of literature. For example, curvaton has
been interpreted in a super-symmetric frame \cite{Postma:2002et,
Enqvist:2003mr} or as a Peccei-Quinn scalar
\cite{Dimopoulos:2003ii}; curvaton mechanism in brane world
picture \cite{Papantonopoulos:2006xi};
the curvaton model constructed by multiple fields has been studied
by Refs. \cite{CG, AVW1, AVW2};
some works have considered various constraints on the
non-Gaussianity of the usual curvaton model, for example, see Ref.
\cite{Huang:2008ze} for the limits from gravitational waves; see
Refs. \cite{Komatsu:2008hk, Beltran:2008ei, Li:2008jn} for the
constraints from iso-curvature perturbations; and also see Ref.
\cite{Ichikawa:2008iq} for the discussion on various possible
cases with large non-Gaussianity.

\section{The DBI-curvaton}

In this section, we study a curvaton model with a non-canonical
kinetic form. Specifically, we consider this model is described by
an DBI action. This action is motivated by string theory
\cite{Aharony:1999ti, Myers:1999ps}\footnote{In Ref.
\cite{Kachru:2003sx}, a system of D- and anti-D-branes in a warped
throat was used to obtain inflation. In recent years, the DBI
action has been applied into brane inflation
models\cite{Dvali:1998pa} and has been studied in
details\cite{Silverstein:2003hf}. Its perturbation theory has been
investigated by Refs. \cite{Garriga:1999vw, Alishahiha:2004eh,
Chen:2005ad}; the non-Gaussianity was studied by Ref.
\cite{Alishahiha:2004eh, Chen:2006nt}; a multiple DBI model was
analyzed by Refs. \cite{Huang:2007hh, Easson:2007dh,
Langlois:2008wt, Langlois:2008qf, Arroja:2008yy}; DBI inflation
can also be discussed by using a one-parameter family of
geometries for the warped throat \cite{Gmeiner:2007uw}.}. So the
scalar field $\sigma$ is interpreted as a moduli parameter of a
D-brane moving in a warped throat. We assume that, the warping
factor and the potential mainly depend on the curvaton field, but
the kinetic parameter is determined by the inflaton field.
Therefore, the effective action is given by
\begin{eqnarray}
{\cal S}_{DBI}=\int d^4x\sqrt{-g}~P(X,\sigma)~,
\end{eqnarray}
and
\begin{eqnarray}
P(X,\sigma) = \frac{1}{f(\sigma)}\left[1-\sqrt{1-2 f(\sigma)
X}\right]-V(\sigma) ~,
\end{eqnarray}
where $f(\sigma)$ is the warping factor. For example,
$f(\sigma)=\lambda/\sigma^4$ in an AdS-like throat. Besides, it is
useful to define a sound speed parameter
\begin{eqnarray}
c_s\equiv\sqrt{1-2 f(\sigma) X}~.
\end{eqnarray}
Notice that, the parameter $X$ is defined as
$X\equiv-\frac{1}{2}g^{\mu\nu}\partial_{\mu}\sigma\partial_{\nu}\sigma-\frac{1}{2}g^{\mu\nu}\partial_{\mu}\phi\partial_{\nu}\phi$
in which $\phi$ plays the role of an inflaton field, so that there
is $\dot\sigma^2\ll\dot\phi^2$ approximately. In FRW universe, the
equation of motion for $\sigma$ is given by
\begin{eqnarray}
\ddot\sigma+3H\dot\sigma-\frac{\dot{c_s}}{c_s}\dot\sigma-c_s
P_{,\sigma}=0
\end{eqnarray}
where $P_{,\sigma}$ denotes the derivative of $P$ with respect to
$\sigma$. The basic scenario of a DBI-curvaton is the same as that
in a usual case, but since the kinetic term is changed, the
spectrum and its non-Gaussianity involves some other parameters,
such as the sound speed $c_s$.

To study the perturbations of the curvaton field, we can split the
scalar field $\sigma$ into
$\sigma(\textbf{x},t)=\sigma_0(t)+\delta\sigma(\textbf{x},t)$,
where $\sigma_0$ represents the spatially homogeneous background
field, and the $\delta\sigma$ stands for the linear fluctuation
which corresponds to the iso-curvature perturbation during
inflation. Taking the gauge on the spatially flat slicing and
using the relation that there is no coupling between entropy and
adiabatic modes, one find that the perturbations in the Fourier
space satisfy the equation of motion as follows,
\begin{eqnarray}
\delta\ddot\sigma_k + \bigg(3- s \bigg)H \delta\dot\sigma_k +
\bigg(\frac{k^2}{a^2}+m_\sigma^2 \bigg) \delta\sigma_k =0~,
\end{eqnarray}
where we have defined $s\equiv\frac{\dot{c_{s}}}{c_{s}H}$,
and the effective mass term $m_\sigma^2$ is given by
\begin{eqnarray}
m_\sigma^2=-c_sP_{,\sigma\sigma}-\frac{P_{,\sigma}^2}{2X}-2c_{s,\sigma}P_{,\sigma}~.
\end{eqnarray}

Finally, we have the perturbation of the curvaton on the hubble
radius
\begin{eqnarray}
\delta\sigma_k=\sqrt{\frac{c_s}{2k^3}} H \bigg( \frac{k}{aH}
\bigg)^{\frac{3}{2}-\nu_\sigma}~,
\end{eqnarray}
with
\begin{eqnarray}\label{nus}
\nu_\sigma^2\simeq\frac{9}{4} + 3\epsilon - \frac{3}{2}s - 3
\eta_\sigma - (1+2c_s^2) \frac{P_{,\sigma}^2}{2XH^2}~.
\end{eqnarray}

In the above formulae, we have ignored the square and the higher
derivative of the slow roll parameters $\epsilon$ and $s$.
Moreover, we have defined a new parameter $\eta_\sigma \equiv -c_s
P_{,\sigma\sigma}/(3H^2)$. The last term of Eq. (\ref{nus}) is
derived by using the assumption that there is no coupling between
adiabatic and iso-curvature modes (the similar analysis has
appeared in Ref. \cite{Langlois:2008mn}).

Further, we can obtain the spectrum of the perturbation
$P_{\sigma}\simeq\frac{c_sH^2}{4\pi^2}$ which gives
$\delta\sigma=\frac{\sqrt{c_s}H}{2\pi}$. To be accurate, the
spectral index of the perturbation is given by
\begin{eqnarray}
n_{\sigma}-1=-2\epsilon+2\eta_{\sigma}+2s+(1+2c_s^2)\frac{P_{,\sigma}^2}{3XH^2}~.
\end{eqnarray}
When considering the non-relative limit $c_s \simeq 1$, the
lagrangian reduces to the canonical kinetic term plus the
potential of the field and so the perturbation spectrum and its
index reduce to the standard form in the usual case.

\section{Generating the Curvature Perturbation}

Now we consider the epoch when inflation has ceased and the
inflaton has already decayed to radiation. During this epoch the
universe is filled with the curvaton field $\rho_\sigma$ and the
radiation $\rho_r$. From this moment on, the curvature
perturbations are generated from the iso-curvature modes because
the pressure perturbation is non-adiabatic. This process ends when
the perturbation becomes adiabatic again, which corresponds to the
epoch of curvaton domination, or that of curvaton decay, whichever
is earlier. The final curvature perturbation can be calculated
when curvaton decays instantaneously $H=\Gamma$. In that case we
can simply consider separately the component curvature
perturbations $\zeta_\sigma$ and $\zeta_r$ on slices of uniform
curvaton density and radiation density. According to the
assumption of curvaton mechanism, the curvature perturbation from
radiation can be neglected, so the total curvature perturbation is
given by
\begin{eqnarray}
\zeta=\frac{3(1+w)\rho_\sigma}{4\rho_r+3(1+w)\rho_\sigma}\zeta_\sigma~.
\end{eqnarray}
Here $w\equiv p_\sigma/\rho_\sigma$ is defined as the
equation-of-state of the curvaton, and $\zeta_\sigma$ is given by
\begin{eqnarray}
\zeta_\sigma=\frac{\delta\rho_\sigma}{3(\rho_\sigma+p_\sigma)}~.
\end{eqnarray}

One can get the linear perturbation of energy density
\begin{eqnarray}
\delta^{(1)}\rho_\sigma &\simeq& \rho_{\sigma,\sigma}\delta\sigma\nonumber\\
&=&
\left[\frac{f_{,\sigma}}{2f^2c_s^3}(1-3c_s^2)+(V-\frac{1}{f})_{,\sigma}\right]\delta\sigma~,
\end{eqnarray}
and to the second order, the perturbation of energy density is
given by
\begin{eqnarray}
\delta^{(2)}\rho_\sigma &\simeq&
\frac{1}{2}\rho_{\sigma,\sigma\sigma}\delta\sigma^2 \nonumber\\
&=& \frac{1}{2}\left[
(\frac{f_{,\sigma\sigma}}{2f^2}-\frac{f_{,\sigma}^2}{f^3})\frac{1-3c_s^2}{c_s^3}+\frac{3f_{,\sigma^2}}{4f^3c_s^5}(1-c_s^2)^2+(V-\frac{1}{f})_{,\sigma\sigma}
\right]\delta\sigma^2~.
\end{eqnarray}

Two possible cases should be considered in the following. First,
if the curvaton dominates the energy density before decays, we
have the final curvature perturbation
\begin{eqnarray}
\zeta\simeq\frac{fc_s}{3(1-c_s^2)}\delta^{(1)}\rho_\sigma~.
\end{eqnarray}
In this case, if we consider that the exit mechanism for the
DBI-curvaton is the process of a tachyon condensate
\cite{Sen:2002nu, Sen:2002in}, we need to take the limit $V\sim
1/f$ \cite{Shiu:2002xp, Cline:2002it}. Correspondingly, one get
the curvature perturbation
\begin{eqnarray}
\zeta\simeq\frac{H_*}{12\pi}\frac{f_{,\sigma}}{fc_s^{3/2}}\frac{1-3c_s^2}{1-c_s^2}~,
\end{eqnarray}
where the result $\delta\sigma=\frac{\sqrt{c_s}H_*}{2\pi}$ has
been used. Moreover, from the definition of $f_{NL}$ (\ref{fnl}),
we have this non-Gaussianity parameter
\begin{eqnarray}
f_{NL}\simeq \frac{5}{2}\left[
\frac{3(1-c_s^2)^3}{(1-3c_s^2)^2}+\frac{4f}{f_{,\sigma}^2}(\frac{f_{,\sigma\sigma}}{2}-\frac{f_{,\sigma}^2}{f})c_s^4\frac{1-c_s^2}{1-3c_s^2}
\right]~,
\end{eqnarray}
under the limit $V\sim 1/f$. Especially, for the tachyon
condensate, there is a matter-like-dominated epoch between the
curvaton decay and the beginning of the hot big bang era
\cite{Chen:2006ni, HenryTye:2006uv}. During this epoch, one has
$c_s\sim0$ which gives the pressureless equation-of-state
$w\sim0$. Thus, one gets the curvature perturbation and the
non-Gaussianity parameter
\begin{eqnarray}
\zeta|_{c_s\sim0} \simeq \frac{H_*}{12\pi}
\frac{f_{,\sigma}}{fc_s^{3/2}}~,~~f_{NL}|_{c_s\sim0} \simeq
\frac{15}{2}~,
\end{eqnarray}
by taking this special limit.

Second, when the curvaton decays, its energy density is some
fraction $r\equiv (\frac{\rho_\sigma}{\rho_r})_D$ of the radiation
density and this fraction is usually very small. So in this case
the curvature perturbation is given by
\begin{eqnarray}
\zeta\simeq\frac{r}{4}\frac{\delta^{(1)}\rho_\sigma}{\rho_\sigma}~.
\end{eqnarray}
Considering the final evolution of the DBI-curvaton is a process
of the tachyon condensate with $V \sim\frac{1}{f}$, one gets the
curvature perturbation
\begin{eqnarray}
\zeta\simeq\frac{rH_*}{16\pi}\frac{f_{,\sigma}}{fc_s^{3/2}}(1-3c_s^2)~,
\end{eqnarray}
and the non-Gaussianity parameter
\begin{eqnarray}
f_{NL}\simeq\frac{10}{3r}\left[
3\frac{(1-c_s^2)^2}{(1-3c_s^2)^2}+\frac{4f}{f_{,\sigma}^2}(\frac{f_{,\sigma\sigma}}{2}-\frac{f_{,\sigma}^2}{f})\frac{c_s^4}{1-3c_s^2}
\right]~,
\end{eqnarray}
respectively. Again, taking the special limit $c_s\sim0$, then
there is
\begin{eqnarray}\label{result}
\zeta|_{c_s\sim0} \simeq
\frac{rH_*}{16\pi}\frac{f_{,\sigma}}{fc_s^{3/2}}~,~~f_{NL}|_{c_s\sim0}
\simeq \frac{10}{r},
\end{eqnarray}
in the DBI-curvaton model.

Furtherly, taking the warping factor $f\simeq
\frac{\lambda}{\sigma^4}$ as an example, we interestingly find
that
\begin{eqnarray}
\zeta=-\frac{rH_*}{4\pi\sigma c_s^{3/2}}~,
\end{eqnarray}
which is a negative value. If assuming $c_s\sim0.01$ and
considering the current observation $|\zeta|\sim10^{-5}$, we can
get the constraint
\begin{eqnarray}
rH_*\sim10^{-7}\sigma~,
\end{eqnarray}
which means the inflation has to take place at a low scale in a
DBI-curvaton model. This scale should be even lower than that
happens in the usual curvaton model. For this reason, in a
DBI-curvaton model it could be hard to detect the primordial
gravitational wave. Moreover, from the observational limit on the
iso-curvature amplitude in the CMB, we learn that the fraction
$r\sim1$ is highly required \cite{Komatsu:2008hk}. Therefore,
there should be $H\sim 10^{-7}\sigma$ and $f_{NL}\sim10$ in a
DBI-curvaton model.

\section{Conclusion and Discussions}

In this paper, we have studied a DBI-curvaton model, in which the
action of curvaton is taken as a DBI type, which
phenomenologically describes a D-brane moving in the AdS throat.
In this case, since the kinetic term is of a non-canonical form,
its perturbations show some different behaviors. For example, the
iso-curvature modes of a DBI action is suppressed by a small
factor $\sqrt{c_s}$. However, after the iso-curvature
perturbations are totally converted into adiabatic ones, the final
curvature perturbation is enhanced when the sound speed is a small
number, as shown in Eq. (\ref{result}). Noting that in
DBI-curvaton $\zeta\sim {1\over c_s^{3/2}}$, which is distinctly
different from that in DBI inflation, in which $\zeta\sim {1\over
c_s^{1/2}}$. In addition, in DBI-curvaton model the
non-Gaussianity parameter $f_{NL}$ is of order $O(10)$, which is
interesting for current obsrvations.
Another interesting issue of the DBI-curvaton model is that, its
exit mechanism is not an oscillating phase, but a process of
tachyon condensate instead. So the final result is related to the
warping factor $f(\sigma)$ but not the potential $V(\sigma)$. This
work in some sense explores a new application of brane-cosmology
in inflation, whose realistic implement in string theory is worth
further study.

\vspace{16mm}

\textbf{Acknowledgments}

We thank Wei Xue and Yi Wang for useful comments on the
manuscript. This work is supported in part by NNSFC under Grant
No: 10775180, in part by the Scientific Research Fund of
GUCAS(NO.055101BM03), in part by CAS under Grant No: KJCX3-SYW-N2.

\end{document}